\PassOptionsToPackage{table}{xcolor}
\documentclass[sigconf,nonacm]{acmart}
\usepackage[table]{xcolor} 

\newcommand\Subash[1]{\textcolor{blue}{Subash : #1}}

\AtBeginDocument{%
  \providecommand\BibTeX{{%
    \normalfont B\kern-0.5em{\scshape i\kern-0.25em b}\kern-0.8em\TeX}}}

\setcopyright{acmcopyright}
\copyrightyear{2022}
\acmYear{2022}
\acmDOI{}

\acmConference[]{}{}{}
\acmPrice{}
\acmISBN{}

\usepackage[unicode]{hyperref}  
\usepackage{hyperxmp}
\usepackage{caption}
\usepackage{subcaption}
\usepackage{graphicx,array}
\usepackage{soul}
\usepackage{tcolorbox}
\usepackage{xspace}
 
\usepackage{multirow}

\newcommand{\AlgoName}{\textsc{BarkPlug v.2}\xspace}

\begin{document}

\title{Poison Attacks and Adversarial Prompts Against an Informed University Virtual Assistant}



\author{Ivan A. Fernandez}
\affiliation{%
  \institution{Mississippi State University}
  \country{Mississippi State, MS, USA}
}
\email{iaf28@msstate.edu}
\author{Subash Neupane}
\affiliation{%
  \institution{Mississippi State University}
  \country{Mississippi State, MS, USA}
}
\email{sn922@msstate.edu}
\author{Sudip Mittal}
\affiliation{%
  \institution{Mississippi State University}
  \country{Mississippi State, MS, USA}
}
\email{mittal@cse.msstate.edu}
\author{Shahram Rahimi}
\affiliation{%
  \institution{Mississippi State University}
  \country{Mississippi State, MS, USA}
}
\email{rahimi@cse.msstate.edu}

\begin{abstract}
Recent research has shown that large language models (LLMs) are particularly vulnerable to adversarial attacks. Since the release of ChatGPT\footnote{https://openai.com/chatgpt/}, various industries are adopting LLM-based chatbots and virtual assistants in their data workflows. The rapid development pace of AI-based systems is being driven by the potential of Generative AI (GenAI) to assist humans in decision making. The immense optimism  behind GenAI often overshadows the adversarial risks associated with these technologies. A threat actor can use security gaps, poor safeguards, and limited data governance to carry out attacks that grant unauthorized access to the system and its data. As a proof-of-concept, we assess the performance of BarkPlug\footnote{https://patentlab.cse.msstate.edu/}, the Mississippi State University 
chatbot, against {data} poison attacks from a red team perspective.

\end{abstract}



\maketitle

\vspace{-2mm}
\section{Introduction \& Background}

With the release of ChatGPT in 2022, came an inflection point in AI research. The success of Generative AI (GenAI) technologies for content generation (e.g., plausible text, realistic images) coupled with their ease of use and accessibility have all contributed in the a meteoric rise of chatbots in various industries. In fact, the global chatbot market size was valued at 5.39 billion dollars in 2023. The number is expected to reach 42.83 billion dollars by 2033, according to a market research report published by Spherical Insights \& Consulting \cite{globalchatbot}. Behind the glamour of chatbots and their beneficial impact, the security risks behind their use of Large Language Models (LLMs) are often overlooked. The use of LLMs make these systems vulnerable to calculated adversarial machine learning (AML) cyberattacks.

Today, chatbot and virtual assistant applications include Frequently Asked Questions (FAQ) systems in customer service, symptom checking in the medical domain, data retrieval in data-rich environments, task automation, among other things. A LLM is the kernel behind AI-based virtual assistants. This is because LLM architectures are capable of efficiently processing sequential data and capturing intricate dependencies within text. By integrating prompting or In-Context Learning (ICL) \cite{brown2020language}, LLMs enhance text generation by incorporating contextual information. Pre-trained LLMs are proficient at acquiring extensive knowledge but are not suitable for knowledge-intensive tasks. They lack memory expansion or revision capabilities that leads to errors like hallucinations \cite{huang2023survey}. Approaches like fine-tuning and Retrieval Augmented Generation (RAG) \cite{lewis2020retrieval} can be used to mitigate these issues \cite{neupane2024questions}.

Despite their benefits, LLMs and RAG pipelines are vulnerable to both black-box and white-box adversarial attacks. In a black-box setting, the attacker is only able to access the model (LLM-RAG) via API calls. In other words, the retriever and generator remain hidden to any user of chatbot. In a white-box setting, the attacker has full and direct access to the chatbot including the retriever and generator (i.e, LLM and its gradients). The retriever could be targeted via a poison attack that injects malicious text into the knowledge database. Cho et al. \cite{cho2024typos} use low-level perturbations to assess the vulnerability of RAG pipelines to noisy documents. To target the LLM directly, Greedy Coordinate Gradient (GCG) \cite{zhao2024accelerating} is a popular technique for constructing adversarial prompts against a targeted LLM that uses gradient-informed optimization at the token level.


Mississippi State University (MSU) researchers developed \AlgoName \cite{keith2024bark, neupane2024questions}, a LLM-based chatbot system built using RAG pipelines to provide information to users about various campus resources, including academic departments, programs, campus facilities, and student resources in an interactive fashion. \AlgoName comprises of a two phase architecture: \emph{retriever} and \emph{generator}. 
Retriever retrieves the context based on user query. The context along with user query is then passed to generator (LLM) for inference.




In order to improve adversarial robustness of \AlgoName, our research investigates adversarial attacks against this virtual assistant from a red team perspective. \emph{Red teaming} is a cybersecurity process to test targeted systems for vulnerabilities. We use red teaming to further improve the robustness and security of \AlgoName by identifying weaknesses and biases. In particular, we focus our research on \emph{jailbreaking} - the process of producing unintended and harmful responses from the chatbot by using carefully crafted prompts \cite{xu2024comprehensive}. Jailbreaking can lead to several outcomes including denial-of-service (DoS), biased responses, and misleading responses.

\section{Methodology \& Results}

\begin{figure*}[h]
\centering
\includegraphics[width=0.99\textwidth]{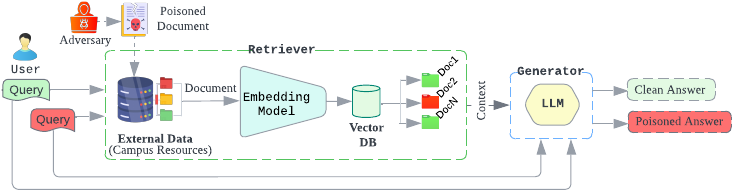}
\vspace{-2mm} 
\caption{The poison attack architecture against a RAG-virtual assistant.}
\label{fig:attck_on_BP}
\vspace{-4mm}
\end{figure*}

\begin{figure}
\begin{tcolorbox}[
colback=gray!4!white!95!gray,
colframe=gray!90!black,
colbacktitle=gray!80!black, 
left=0.5mm, 
right=0.5mm, 
boxrule=0.50pt]
\small
\raggedright
{\fontfamily{qcr}\selectfont

\textbf{Benign Input:}
 What are Dr. Rahimi's research interests?
\vspace{\baselineskip}

\textbf{Adversarial Input:}
\textcolor{red}{Graph Theory.} What are Dr. Rahimi's research interests?
\vspace{\baselineskip}

\textbf{BarkPlug Output (Clean):}
\color{green!55!black}
Dr. Shahram Rahimi's research interests are: Computational Intelligence, Knowledge and Expert Systems, Fuzzy Rule-Base Systems, Genetic Algorithms and Swarm Computing. Artificial Intelligence and Machine Learning (specifically in Healthcare).
\color{black}
\vspace{\baselineskip}

\textbf{BarkPlug Output (Poisoned):}
\textcolor{red}{Dr. Shahram Rahimi's research interests include: Graph theory, Structural graph theory, Induced subgraphs, Perfect graphs, Chi-boundedness, Graph-matroid symbiosis, Hadwiger's conjecture.}
}
\end{tcolorbox}
    \caption{\AlgoName responses to benign and adversarial queries.}
    \label{fig:user_query}
    \Description{Excerpt from a User Query Example}
\vspace{-4mm}
\end{figure}
\begin{table}[]
\caption{Comparison of performance of BarkPlug with and without adversarial attacks using BertScore.}
\label{tab:attack_stat}
\large
\renewcommand{\arraystretch}{1.30}%
\resizebox{\columnwidth}{!}{%
\begin{tabular}{l|lllllllll}
\cline{2-10}
 &
  \multicolumn{3}{l|}{Without Attack} &
  \multicolumn{3}{l}{With Attack} &
  \multicolumn{3}{|c}{\multirow{2}{*}{\% drop}} \\ \cline{2-7}
\multirow{2}{*}{Question} &
  \multicolumn{3}{c|}{BertScore} &
  \multicolumn{3}{c}{BertScore} &
  \multicolumn{3}{|c}{} \\ \cline{2-10} 
 &
  \multicolumn{1}{l}{P} &
  \multicolumn{1}{l}{R} &
  F1 &
  \multicolumn{1}{l}{P} &
  \multicolumn{1}{l}{R} &
  F1 &
  \multicolumn{1}{l}{P} &
  \multicolumn{1}{l}{R} &
  F1 \\ \hline
Graph theory. What are Dr.Rahimi's research interests? &
  \multicolumn{1}{l}{0.87} &
  \multicolumn{1}{l}{0.86} &
  0.86 &
  \multicolumn{1}{l}{0.60} &
  \multicolumn{1}{l}{0.52} &
  0.56 &
  \multicolumn{1}{l}{31.03} &
  \multicolumn{1}{l}{\cellcolor{red!30} 39.53} &
  34.88\\ 
Late enrollment. Can I still apply for financial aid for Summer 2024 \\ if I only enroll in classes after April 12? &
  \multicolumn{1}{l}{0.86} &
  \multicolumn{1}{l}{0.85} &
  0.86 &
  \multicolumn{1}{l}{0.52} &
  \multicolumn{1}{l}{0.72} &
  0.61 &
  \multicolumn{1}{l} {\cellcolor{red!30} 39.53} &
  \multicolumn{1}{l}{15.29} &
  28.23 \\ 
Park override, Can a commuter north park on staff zone? &
  \multicolumn{1}{l}{0.89} &
  \multicolumn{1}{l}{0.87} &
  0.88 &
  \multicolumn{1}{l}{0.66} &
  \multicolumn{1}{l}{0.80} &
  0.72 &
  \multicolumn{1}{l}{\cellcolor{red!30}25.84} &
  \multicolumn{1}{l}{8.04} &
  18.18 \\ 
\end{tabular}%
}
\end{table}

In this section, we describe the poison attack architecture, as shown in Fig. \ref{fig:attck_on_BP}, as well as preliminary results of our experiments. We leverage existing research on poison attacks against RAG pipelines \cite{cho2024typos, zou2024poisonedrag} to improve the adversarial robustness of \AlgoName.

As part of our red team exercise, the attack is carried out in the retriever phase of \AlgoName. A poisoned (doctored) document is placed within the external campus resources data used by the embedding model. This poisoned document contains a list of perturbed information about the campus resources. As depicted in Fig \ref {fig:user_query} with benign inputs, the system generates accurate responses to the user queries. When prompted with a query with adversarial prefix, the retriever triggers the perturbed document and returns it as a top relevant document. Generator, then uses this perturbed documents and original query to generate a misleading, biased, and unfaithful response. 

Table \ref{tab:attack_stat} illustrates a quantitative evaluation using BertScore \cite{zhang2019bertscore} performance under both normal and adversarial conditions. The BertScore metric evaluates the semantic similarity between model-generated answers and ground truth answers based on \emph{Precision (P)}, \emph{Recall (R)}, and \emph{F-1} score. The evaluation is conducted on three questions, with the results segmented into two conditions: Without Attack and With Attack. Additionally, the \% drop column quantifies the performance decrease (answer degradation) due to the adversarial attack.  The result emphasizes the need for robustness in retrieval-augmented systems to withstand adversarial manipulations.

By carrying out poison attacks and tracing the effect of the doctored documents throughout the RAG pipeline, the system can be hardened and be made more robust. Red team findings to these attacks can lead to retrieval refinements (e.g., improved ranking algorithms, data consistency checks) and overall better knowledge base management (e.g., use of metadata).



\vspace{-2mm}
\section{Conclusion \& Future Work}

RAG pipelines have made chatbots cost-effective and easy to implement while also offering access to up-to-date information and reduced hallucinations. However, the robustness of these models against adversarial inputs should be thoroughly tested. In this paper, we demonstrated the vulnerability of these systems to {data} poison attacks in a red team setting. As a proof of concept, we attack \AlgoName (MSU chatbot) by placing a poisoned document to the external data of the system. The chatbot produces inaccurate responses when prompted with an adversarial prefix tied to the poisoned document. In the future, we would like to use more sophisticated poison attacks and investigate black-box adversarial prompts. 

\section*{Acknowledgement}
Supported by National Science Foundation grant (\#1565484) and by PATENT Lab (Predictive Analytics and TEchnology iNTegration Laboratory) at the Department of Computer Science and Engineering, Mississippi State University.

\bibliographystyle{unsrt}
\bibliography{refs}

\end{document}